# Fiber-laser-pumped Ti:sapphire laser

G. K. Samanta,[1,*] S. Chaitanya Kumar,[1] Kavita Devi,[1] and M. Ebrahim-Zadeh[1,2]

[1]*ICFO-Institut de Ciencies Fotoniques, Mediterranean Technology Park, 08860 Castelldefels, Barcelona, Spain*

[2]*Institucio Catalana de Recerca i Estudis Avancats (ICREA), Passeig Lluis Companys 23, Barcelona 08010, Spain*

[*] *Email: gsamanta@gmail.com  Fax: 34 935534000*

# ABSTRACT

We report the first experimental demonstration of efficient and high-power operation of a Ti:sapphire laser pumped by a simple, compact, continuous-wave (cw) fiber-laser-based green source. The pump radiation is obtained by direct single-pass second-harmonic-generation (SHG) of a 33-W, cw Yb-fiber laser in 30-mm-long MgO:sPPLT crystal, providing 11 W of single-frequency green power at 532 nm in TEM$_{00}$ spatial profile with power and frequency stability better than 3.3% and 32 MHz, respectively, over one hour. The Ti:sapphire laser is continuously tunable across 743-970 nm and can deliver an output power up to 2.7 W with a slope efficiency as high as 32.8% under optimum output coupling of 20%. The laser output has a TEM$_{00}$ spatial profile with $M^2$<1.44 across the tuning range and exhibits a peak-to-peak power fluctuation below 5.1% over 1 hour.

# 1. Introduction

Titanium-doped sapphire (Ti:sapphire) is the most successful solid-state laser material in the near-infrared wavelength range due to its high saturation energy, large stimulated emission cross-section, and broad absorption gain bandwidths [1]. It has been extensively developed for continuous-wave (cw) operation, ultra-short pulse generation, high-power amplification, and much more, and has been successfully deployed in a wide range of applications from high-intensity physics, frequency metrology, spectroscopy, as well as pumping of tunable optical parametric oscillators (OPOs). Although Ti:sapphire has broad absorption bandwidth, due to the relatively weak absorption peak in the blue-green wavelength range, its successful operation requires high-power blue-green pump sources. As such, Ti:sapphire lasers have been pumped with multiwatt argon-ion [1,2], copper-vapor [3], and most notably frequency-doubled all-solid-state green lasers [1,4], resulting in fairly bulky, complicated and expensive setups. For further progress in Ti:sapphire laser technology, it would be desirable to devise more simplified pump laser designs to reduce system complexity and cost, while maintaining or enhancing device performance with regard to all important operating parameters. Recently, a cw Ti:sapphire laser pumped directly by a GaN diode laser in the blue was reported [5], but the limited pump powers available from diode lasers in good spatial beam quality restrict the effectiveness of this approach only to low-power cw operation. On the other hand, optically-pumped-semiconductor-lasers (OPSL) in the green can in principle be used to pump cw Ti:sapphire laser [6], but limited progress has been achieved in this area so far, leaving open the need for the development of powerful alternative green sources with high spatial quality and in simple, practical all-solid-state design to pump high-power cw or mode-locked Ti:sapphire lasers.

A major step in this direction is the development of novel green sources based on second-harmonic-generation (SHG) of the rapidly advancing fiber laser technology to replace the relatively complex and expensive cw solid-state green lasers at 532 nm. Among all existing frequency-doubling techniques, such as intracavity, multipass or resonant enhancement, external single-pass SHG is the most direct route for the realization of a compact, cost-effective and practical green source in simple design. Moreover, single-pass SHG avoids the traditional "green problem" [7], commonly observed in intracavity doubling of solid-state lasers. At the same time, single-pass SHG of fiber lasers offers additional advantages of air-cooling and flexible packaging, making the fiber pumping approach highly attractive over solid-state green sources. Recently, we demonstrated an efficient and high-power green source based on direct single-pass SHG of a cw Yb-fiber laser in MgO-doped stoichiometric periodically-poled lithium tantalate (MgO:sPPLT) crystal, providing up to 9.6 W of power at 532 nm in $TEM_{00}$ spatial beam with power stability of 9% over 13 hours [8,9]. As such, we were able to successfully deploy this source to pump a high-power cw OPO for the near-infrared [10].

Here, we report the successful use of this simple, high-power, cw, fiber-based green source to pump a Ti:sapphire laser, generating an output power of >2.7 W with tunability of >227 nm in a $TEM_{00}$ spatial profile. Although operation of a Ti:sapphire laser was previously reported using a pulsed frequency-doubled fiber laser [11], to our knowledge, this is the first report of a Ti:sapphire laser pumped by a cw fiber laser source, which also demonstrates the competitive performance of the source compared to the well-established solid-state green pump lasers at 532 nm.

## 2. Experimental set-up

The schematic of the experimental setup is shown in Fig. 1. The fundamental pump source is a cw, single-frequency Yb-fiber laser (IPG Photonics, YLR-30-1064-LP-SF) at 1064 nm, providing a linearly polarized output beam with $M^2$<1.01 and a nominal linewidth of 89 kHz. A 30-mm-long MgO:sPPLT crystal [10], containing a single grating (Λ=7.97 µm), is used for single-pass SHG into the green at 532 nm. At the highest available fundamental power of ~33 W, our SHG source provides as much as 11 W of green power in a TEM$_{00}$ spatial profile ($M^2$<1.3) with a frequency stability better than 32 MHz over 1 hour. However, since our earlier reports [8,9], we have improved the output power stability of the green source from 7.9% to better than 3.3% peak-to-peak, measured over 1 hour. This improvement has been obtained by using a novel uniform crystal heating configuration, where the crystal is placed at the center of a large-area oven used to heat the 50-mm-long crystal. As compared to the small-area oven used previously [8,9], the central part of the new oven provides higher temperature uniformity and is less sensitive to the instability (±0.1 °C) of the temperature controller. As a result, it maintains improved temperature stability over the full crystal length, resulting in higher green output power stability. The fiber green source is so robust that it takes only 30 minutes to reach stable output power during the fiber laser warm-up time. We have operated the source reliably and regularly on a daily basis for over two years, without any degradation in output power or stability.

In the present experiment, to maintain stable output characteristics, we operated our green source at maximum power and used an attenuator comprising a half-wave-plate (HWP) and a polarizing beam-splitter (PBS) to vary the input power to the Ti:sapphire laser. Using plano-convex lens (L$_2$), we focused the green pump beam to different waists at the center of the 10-mm-long, Brewster-cut Ti:sapphire crystal (0.15 wt.% doping, *FOM*>270), which is located on brass slab, and water-cooled only on the lower side. The green beam was polarized along the *c*-axis of the crystal to maximize absorption [1,2], which we measured to be >80%. The laser was configured in an astigmatic-compensated, standing-wave cavity, comprising two concave mirrors, $M_1$ and $M_2$, and a plane output coupler (OC). To access the wide tuning range of the Ti:sapphire laser, we used two sets of concave mirrors, all of the same curvature (*r*=10 cm), providing high reflectivity (>99%) across 760-840 nm and 840-1000 nm. For wavelength tuning and control, we used a birefringent filter (BRF).

## 3. Results and discussion

### 3.1. Output power across the tuning range

We investigated the output power of the Ti:sapphire laser across the tuning range using both sets of concave mirrors, by deploying different output couplers of varying transmission across 740-970 nm. Tuning was achieved using the BRF. The results are shown in Fig. 2, where the output power across the obtained tuning range is plotted for three different output couplers. For OC1 (*T*=5%-0.74% over 780-970 nm), the output power exactly follows the transmission curve (inset of Fig. 2) up to the wavelength ~940 nm, with no evidence of the expected drop in output power due to gain reduction in the Ti:sapphire crystal at longer wavelengths [1]. This can be attributed to the high intracavity power due to the combination of high pump power (10.5 W) and low output coupling, pointing to the possibility of extracting higher power across the tuning range using larger output coupling. However, beyond ~940 nm, there is a drop in output power with the increase in output coupling, due to the lower gain of the Ti:sapphire crystal.[1] Tuning below 780 nm is limited by the larger loss due to increasing output coupling at shorter wavelengths. Therefore, using OC1 and both sets of cavity mirrors, the Ti:sapphire laser can be continuously tuned over 780-970 nm with a maximum power of 1 W at 780 nm.

To enhance the tuning range and output power, we deployed two additional output couplers of varying transmission, OC2 ($T$=4.4%-3.12% across 743-800 nm) and OC3 of ($T$=25%-18.6% across 767-812 nm), also shown in the inset of Fig. 2. With OC2, we obtain tuning across 743-800 nm with a maximum power of 0.85 W at 795 nm. With OC3, we obtain an output power up to 2.04 W at 795 nm for an output coupling of 20%, with >1.12 W available across 767-812 nm. Hence, using two sets of cavity mirrors and three output couplers, we achieve a total wavelength tuning range of ~227 nm across 743-970 nm and generate up to 2.04 W of output power. Wavelength tuning below 743 nm is limited by the reflectivity fall-off of the cavity mirrors, whereas coverage beyond 970 nm was limited by the free spectral range of the BRF.

## 3.2. Optimization of output coupling

To optimize the output coupling for maximum power extraction, we operated the Ti:sapphire laser in free-running cavity at a fixed pump power of ~10.5 W (at the Ti:sapphire crystal) and used several additional output couplers with transmission ranging from 1.4% to 40.7% at ~812 nm, where the laser operated due to the lowest cavity loss in the absence of BRF. We measured the extracted power and corresponding operation threshold of the Ti:sapphire laser as a function of output coupling, with the results shown in Fig. 3. The output power rises from 1 W to 2.24 W with the increase in output coupling from 1.4% to 19.9%. However, further increases in output coupling result in reduced output power with a clear peak near 20%, implying an optimum output coupling of 20% for our laser. As expected, the laser threshold increases linearly with the output coupling and reaches 6 W at 40.7%, with an output power of 1.66 W. Higher output coupling is possible with our laser for the available pump power, but will result in reduced output power. It is to be noted that the temperature rise in the Ti:sapphire crystal due to pump absorption effectively decreases the fluorescence lifetime and quantum efficiency by ~10% [12] and thermo-optical aberrations due to temperature gradients in the crystal also limit the laser output power [13]. Therefore, an enhancement in output power is feasible by implementing further thermal management of the crystal by water cooling from four sides.

### 3.3. Power scaling

Having determined the optimum output coupling of 20%, we investigated power scaling of the Ti:sapphire laser. Figure 4 (a) shows the variation of output power as the function of pump power without the BRF, for two different pump beam waist radii of 19 μm and 30 μm. With a pump beam waist of 19 μm and 20% output-coupling, the Ti:sapphire laser has a threshold of 2.51 W and the output power increases with pump power with a slope efficiency of 31.8%, reaching a maximum value of 2.7 W at ~11 W of pump power. With a pump beam waist of 30 μm, the laser threshold increased to 3.27 W, but the output power rises with a slope efficiency as high as 32.9%, reaching 2.22 W for 10.57 W of pump power. It may, thus, be concluded that the laser operates robustly with an output power >2.2 W and slope efficiency >31.8% for a range of focused pump beam waists from 19 μm to 30 μm. Further enhancement in output power is also expected by using tighter pump focusing. Using the BRF, we also measured the power scaling of the Ti:sapphire laser at two different wavelengths of 795 nm and 867 nm for the same pump beam waist of 30 μm, as shown in Fig. 4 (b). At 795 nm, the laser has a threshold of 3.33 W and generates an output power as high as 2.02 W with a slope efficiency of 28%. On the other hand, at 867 nm we have a maximum output power of 1.1 W, a slope efficiency of 21.8%, and higher threshold of 4 W, due to gain reduction in the Ti:sapphire crystal away from the emission peak at 795 nm [1].

### 3.4. Power stability

We also recorded the power stability of the green pump laser and the Ti:sapphire laser at the maximum output power of 11 W and 2.7 W, respectively, with the results shown in Fig. 5. As evident from Fig. 5 (a), without any thermal isolation and control, the green source has a better peak-to-peak power fluctuation (<3.3% over 60 minutes) than our previous reports [8,9]. Similarly, Fig. 5(b) shows the corresponding peak-to-peak power fluctuation of the Ti:sapphire laser to be <5.1% at 812 nm over 60 minutes. The higher power fluctuation of the Ti:sapphire laser can be attributed to factors such as thermal effects in Ti:sapphire rod, mechanical instabilities in the present setup, and contributions from the green pump power instability. We expect higher stability through improvements in the green pump power stability, as well as better mechanical and thermal isolation of both the green pump and the Ti:sapphire laser from the laboratory environment.

### 3.5. Beam quality

The typical far-field energy distribution of the Ti:sapphire laser at 823 nm, together with the intensity profile and Gaussian fits in the two orthogonal directions, measured at a distance >3 m away from the output coupler, is shown in Fig. 6. Although the profile appears to confirm a Gaussian distribution with an ellipticity >88%, using a focusing lens ($f$=25 cm) and scanning beam profiler we measured the $M^2$ values of the output beam at two different output powers obtained with two different output couplers, OC1 and OC3, to confirm $TEM_{00}$ spatial mode. For OC1, the $M^2$-values of the output beam at >200 mW power at 823 are measured as $M_x^2$<1.27 and $M_y^2$ <1.21, while for OC3, the $M^2$-values of the output beam at 2.02 W power at 795 nm are measured as $M_x^2$<1.3 and $M_y^2$ <1.44, thus confirming TEM$_{00}$ spatial mode. Considering the accuracy (±5%) of the $M^2$-measurement system, the Ti:sapphire output beam has almost similar $M^2$-values at different power levels throughout the tuning range. Although the laser output has slightly higher $M^2$-value than the diffraction limit, the beam quality can be further improved using proper thermal management of the Ti:sapphire rod with efficient water cooling.

## 4. Conclusions

In conclusion, we have demonstrated the first operation of a Ti:sapphire laser pumped by a cw fiber laser green source. We have achieved cw laser output over a wide tuning range (227 nm) across 743-970 nm. A maximum output power of 2.7 W and a slope efficiency as high as 32.8% has been obtained at 812 nm, with optimized output coupling of 20%. The Ti:sapphire laser output exhibits TEM$_{00}$ spatial quality ($M_x^2$<1.3 and $M_y^2$ <1.44), with a power fluctuation below 5.1%, which can further be improved by enhancing the power stability of the green source with mechanical and thermal isolation. The overall performance of the device is competitive with the well-established Ti:sapphire lasers pumped by diode-pumped solid-state lasers at 532 nm, confirming the practical viability of fiber-laser-pumping, while greatly reducing system complexity and cost together with the advantages of air-cooling and flexible packaging. Further enhancement in the output power and wavelength tuning to cover the entire emission spectrum of the Ti:sapphire are possible using more efficient cooling of the Ti:sapphire crystal, tighter pump focusing, and optimized broadband mirrors.

## 5. Acknowledgements

This work was supported by the Ministry of Innovation and Science, Spain, through grant Novalight (TEC2009-07991) and the Consolider project, SAUUL (CSD2007-00013). We also acknowledge partial support by the European Union 7$^{th}$ Framework Program, MIRSURG (224042).

**References**

1. P. F. Moulton, J. Opt. Soc. Am. B **3**, 125 (1986)

2. P. A. Schulz, IEEE J .Quantum Electron. **24**, 1039 (1988); C. Zimmermann, V. Vuletic, A. Hemmerich, L. Ricci, and T. W. Hansch, Opt. Lett. **20**, 297 (1995)

3. M. R. H. Knowles and C. E. Webb, Opt. Lett. **18**, 607 (1993)

4. J. Harrison, A. Finch, D. M. Rines, G. A. Rines, and P. F. Moulton, Opt. Lett. **16**, 581 (1991)

5. P. W. Roth, A. J. Maclean, D. Burns, and A. J. Kemp, Opt. Lett. **34**, 3334 (2009)

6. B. Resan, E. Coadou, S. Petersen, A. Thomas, P. Walther, R. Viselga, J. M. Heritier, J. Chilla, W. Tulloch, and A. Fry, Solid State Lasers XVII: Technology and Devices, SPIE **6871**, 87116 (2008)

7. T. Baer, J. Opt. Soc. Am. B 3, 1175 (1986)

8. G. K. Samanta, S. Chaitanya Kumar, and M. Ebrahim-Zadeh, Opt. Lett., 34, 1561 (2009)

9. S. Chaitanya Kumar, G. K. Samanta, and M. Ebrahim-Zadeh, Opt. Express 17, 13711 (2009)

10. G. K. Samanta, S. Chaitanya Kumar, R. Das, and M. Ebrahim-Zadeh, Opt. Lett. 34, 2255 (2009)

11. Hsiao-hua Liu, CLEO, JTuD3, (2009)

12. P. Albers, E. Stark, and G. Huber, J. Opt. Soc. Am. B 3, 134 (1986)

13. P. A. Schulz and S. R. Henion, IEEE J. Quantum Electron. 27, 1039 (1991)

**Figure Captions**

**Figure 1.**

Fig.1. Schematic of the fiber-laser-green-pumped cw Ti:sapphire laser. λ/2, half-wave plate; PBS, polarizing beam splitter; L, lens; M, mirrors; OC, output coupler.

**Figure 2.**

Fig.2. Extracted output power across the tuning range of the Ti:sapphire laser using three different output couplers and two set of cavity mirrors. Inset: transmission of the output couplers OC1, OC2 and OC3 versus wavelength.

**Figure 3.**

Fig.3. Dependence of the extracted output power and threshold power of the Ti:sapphire laser on output coupling at ~812 nm . Lines are to guide the eye.

**Figure 4.**

Fig.4. Variation of the output power of the Ti:sapphire laser with input power (a) for different pump waists, and (b) at different wavelengths. Output coupling is 20%.

**Figure 5.**

Fig.5. Time-trace of the (a) frequency-doubled green pump power and (b) Ti:sapphire output power, at 11 W and 2.7 W respectively over 1 hour.

**Figure 6.**

Fig. 6. Far-field $TEM_{00}$ energy distribution and intensity profiles of the Ti:sapphire output beam at 823 nm.

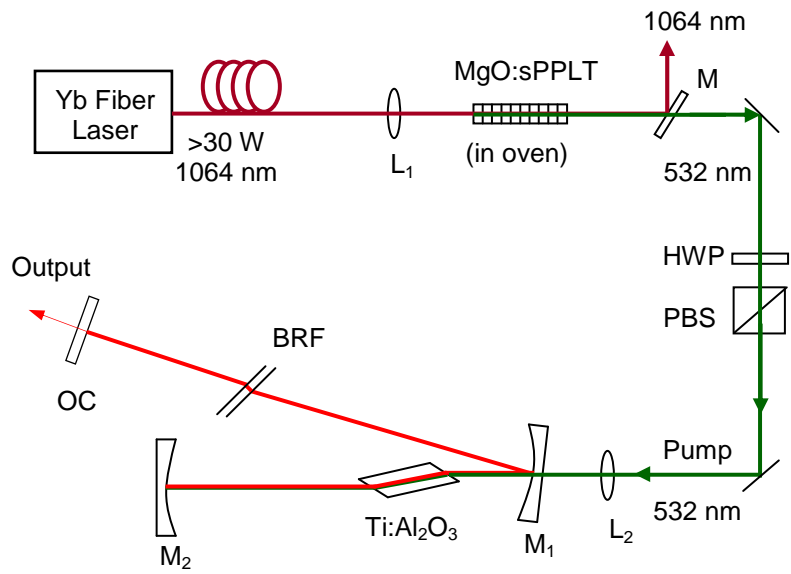

**Figure 1**

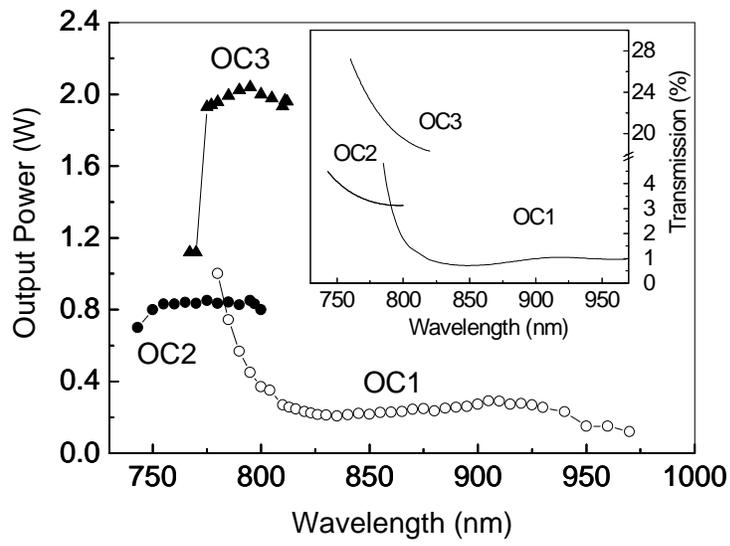

**Figure 2**

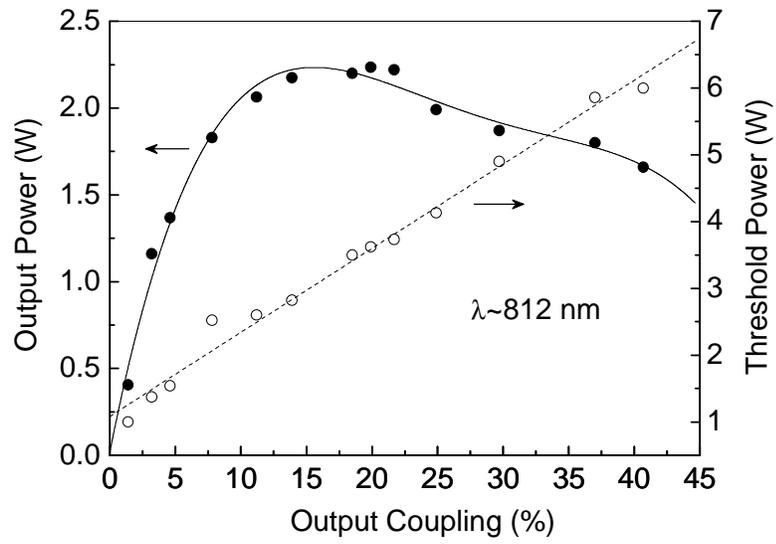

**Figure 3**

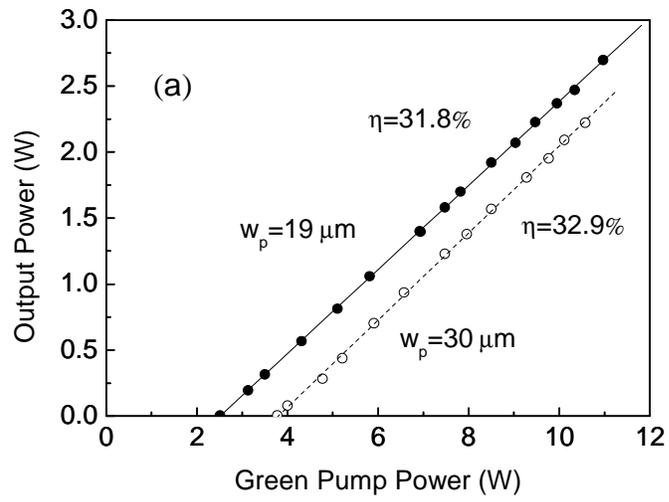
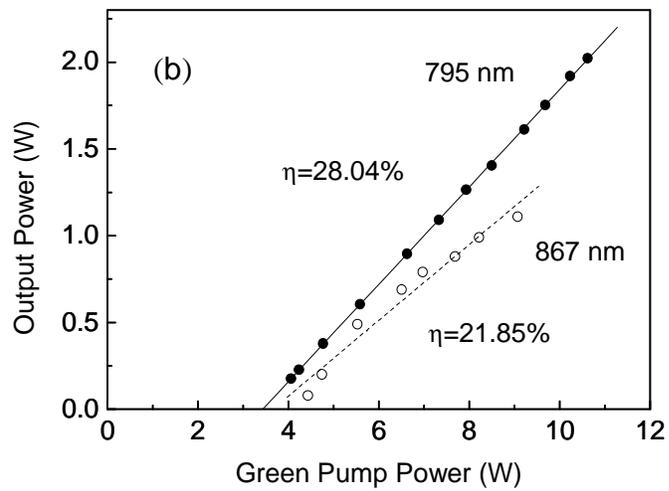

**Figure 4**

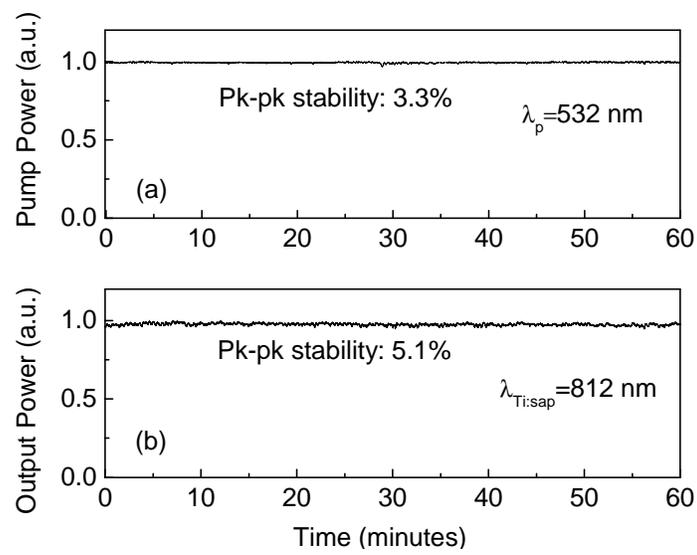

**Figure 5**

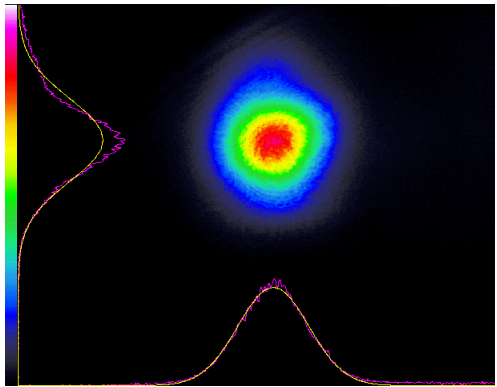

**Figure 6**